\documentclass[reprint,showpacs,noeprint,amsmath,a4paper,amssymb,xcolor=dvipsnames,superscriptaddress,floatfix]{revtex4-1}
\pdfoutput=1

\usepackage[T1]{fontenc}
\usepackage[pdftex]{graphicx}% Include figure files
\usepackage{bm}% bold math
\usepackage[pdftex,colorlinks=true,citecolor=blue,urlcolor=blue,linkcolor=black]{hyperref}
\usepackage{braket}
\usepackage[dvipsnames]{xcolor}
\usepackage[version-1-compatibility,binary-units,abbreviations=true]{siunitx}
\usepackage{comment}

\usepackage{physics}
\usepackage{mathrsfs}

\begin{document}
%\bibliographyunit[\chapter]
%\begin{bibunit}

%%%%%%%%%%%%%%%%%%%%%%%%%%%%%%%%%%%%%%%%%%%%%%%%%%%%%%%%%%%%%%%%%%%%%%%%%%%%%%%%%
%						    	Title and Authors	    						%
%%%%%%%%%%%%%%%%%%%%%%%%%%%%%%%%%%%%%%%%%%%%%%%%%%%%%%%%%%%%%%%%%%%%%%%%%%%%%%%%%

\title{\textcolor{black}{Emergent interaction-driven elliptic flow of few fermionic atoms}}
\author{Sandra~Brandstetter}
    \thanks{These authors contributed equally to this work.}
	\affiliation{Physikalisches Institut der Universit\"at Heidelberg, Im Neuenheimer Feld 226, 69120 Heidelberg, Germany} 
     
\author{Philipp~Lunt}
    \thanks{These authors contributed equally to this work.}
	\affiliation{Physikalisches Institut der Universit\"at Heidelberg, Im Neuenheimer Feld 226, 69120 Heidelberg, Germany}

 \author{Carl~Heintze}
	\affiliation{Physikalisches Institut der Universit\"at Heidelberg, Im Neuenheimer Feld 226, 69120 Heidelberg, Germany}
 \author{Giuliano~Giacalone}
	\affiliation{Institut f\"{u}r Theoretische Physik, Universit\"{a}t Heidelberg, Philosophenweg 16, 69120 Heidelberg, Germany}
 \author{Lars~H.~Heyen}
	\affiliation{Institut f\"{u}r Theoretische Physik, Universit\"{a}t Heidelberg, Philosophenweg 16, 69120 Heidelberg, Germany}
 \author{ Maciej~Ga\l ka}
	\affiliation{Physikalisches Institut der Universit\"at Heidelberg, Im Neuenheimer Feld 226, 69120 Heidelberg, Germany}
 \author{ Keerthan~Subramanian}
	\affiliation{Physikalisches Institut der Universit\"at Heidelberg, Im Neuenheimer Feld 226, 69120 Heidelberg, Germany}
 \author{Marvin~Holten}
    
	\affiliation{Physikalisches Institut der Universit\"at Heidelberg, Im Neuenheimer Feld 226, 69120 Heidelberg, Germany}
    \affiliation{Current adress: Vienna Center for Quantum Science and Technology, Atominstitut, TU Wien, Vienna, Austria}
   
\author{Philipp~M.~Preiss}
	\affiliation{Physikalisches Institut der Universit\"at Heidelberg, Im Neuenheimer Feld 226, 69120 Heidelberg, Germany}
    \affiliation{Current adress: Max Planck Institute of Quantum Optics, Hans-Kopfermann-Str. 1, 85748 Garching, Germany}
 \author{Stefan~Floerchinger}
	\affiliation{Theoretisch-Physikalisches Institut, Friedrich-Schiller-Universität Jena, Max-Wien-Platz 1, 07743 Jena, Germany}
\author{Selim~Jochim}
	\affiliation{Physikalisches Institut der Universit\"at Heidelberg, Im Neuenheimer Feld 226, 69120 Heidelberg, Germany}

\date{\today}

\maketitle

%%%%%%%%%%%%%%%%%%%%%%%%%%%%%%%%%%%%%%%%%%%%%%%%%%%%%%%%%%%%%%%%%%%%%%%%%%%%%%%%%
%						        	Abstract        	   						%
%%%%%%%%%%%%%%%%%%%%%%%%%%%%%%%%%%%%%%%%%%%%%%%%%%%%%%%%%%%%%%%%%%%%%%%%%%%%%%%%%

\textbf{Hydrodynamics provides a successful framework to effectively describe the dynamics of complex many-body systems ranging from subnuclear to cosmological scales by introducing macroscopic quantities such as particle densities and fluid velocities. According to textbook knowledge, it requires coarse graining over microscopic constituents to define a macroscopic fluid cell, which is large compared to the interparticle spacing and  the mean free path. In addition, the entire system must consist of many such fluid cells.  \textcolor{black}{In high energy heavy ion collisions, hydrodynamic behaviour is inferred from the observation of elliptic flow.}
Here, we demonstrate the emergence of  \textcolor{black}{elliptic flow} in a system of few strongly interacting atoms. In our system a hydrodynamic description is a priori not applicable, as all relevant length scales, i.e. the system size, the inter-particle spacing, and the mean free path are comparable. 
The single particle resolution, deterministic control over particle number and interaction strength in our experiment allow us to explore the boundaries between a microscopic description and a hydrodynamic framework in unprecedented detail.}

%%%%%%%%%%%%%%%%%%%%%%%%%%%%%%%%%%%%%%%%%%%%%%%%%%%%%%%%%%%%%%%%%%%%%%%%%%%%%%%%%
%						        	Main Text         	   						%
%%%%%%%%%%%%%%%%%%%%%%%%%%%%%%%%%%%%%%%%%%%%%%%%%%%%%%%%%%%%%%%%%%%%%%%%%%%%%%%%%

%Hydrodynamics is one of the basic principles governing the behavior of matter.  The anisotropy of pressure-gradient forces leads to faster acceleration along the direction of tighter confinement. As a consequence, the expanding fluid undergoes a characteristic inversion of its initial aspect ratio. 

\textcolor{black}{Elliptic flow \cite{Ollitrault_1992} is the characteristic inversion of the initial aspect ratio, arising from the anisotropic pressure-gradient forces of a system with an initially elliptic geometry.} In the last decades, it has been used to probe the hydrodynamic behaviour of the quark-gluon plasma ~\cite{Braun-Munzinger_2007,Busza:2018rrf} in the context of relativistic nuclear collisions, where it is inferred from an elliptical deformation of the momentum distribution of the detected particles \cite{Poskanzer_1998,Bhalerao:2005mm}. 
In addition, it has been observed at vastly different energy scales with expanding ultracold quantum gases \cite{Thomas_2002,Davis_1995, Trenkwalder_2011, Fletcher_2018} \textcolor{black}{where the characteristic inversion of the initial aspect ratio can be measured directly}. 
More recently, elliptic flow has been seen in proton-nucleus \cite{ALICE:2012eyl,ATLAS:2012cix,PHENIX:2018lia,STAR:2022pfn} and even proton-proton collisions \cite{Khachatryan_2010} producing few tens of final state particles \cite{Nagle:2018nvi,Schenke:2021mxx}. This indicates emergent collective behaviour of matter in extreme conditions where standard criteria for the applicability of hydrodynamics do not hold \cite{Kurkela:2019kip,Ambrus:2022qya}.

In this work, we study the quantum dynamics of few contact-interacting fermionic atoms after release from an elliptically shaped trap. We measure the position or momentum of each atom after different expansion times. For a system of as few as ten constituents, the initial real space density is shaped elliptically while the momentum space distribution is isotropic. During the expansion, we observe a redistribution of momenta due to interactions, leading to the characteristic inversion of the initial aspect ratio. This redistribution sets it apart from the single particle case, where the inversion of the aspect ratio is caused by the anisotropy of the initial momentum distribution. Our control over the atom number and interaction strength allows us to explore the emergence of elliptic flow from the single particle limit.

 \begin{figure*}
    \centering
	\includegraphics{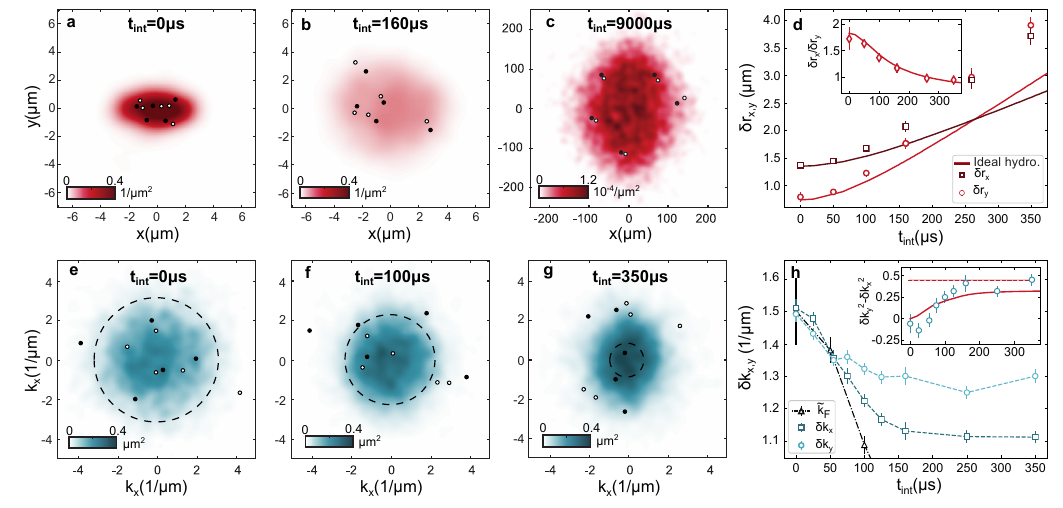}
    \caption{
    \textcolor{black}{\textbf{Elliptic flow of ten fermions.}} We prepare 5+5 strongly interacting spin up and down atoms (black/white dots) in the ground state of an elliptically shaped trap. We measure their positions (\textbf{a-c}) or momenta (\textbf{e-g}). The two dimensional histograms show the density distribution, obtained from averaging over many experimental realizations of the same quantum state.
    The initial system has an elliptic density distribution in real space and a round Fermi surface in momentum space (see \textbf{a} and \textbf{e}). 
    We study the expansion after switching off the trap (\textbf{b-c}, \textbf{f-g}) and observe the inversion of the \textcolor{black}{initial} aspect ratio in real space and the build up of momentum anisotropy. 
    The dashed black circle in \textbf{e-g} shows the Fermi momentum calculated from the real space peak density. 
    \textbf{d} Root mean square of the atom positions $\delta r_\text{x}, \delta r_\text{y}$ as a function of $t_\text{int}$. The expectation assuming ideal hydrodynamic evolution of the corresponding many-body system with the same initial density is shown as a reference (red lines). \textcolor{black}{The inset shows the anisotropy $\delta r _\text{x}/\delta r_\text{y}$ of the expanding density. The red line corresponds to the ideal hydrodynamic expansion.}
    \textbf{h} Root mean square value $\delta k_\text{x},\delta k_\text{y}$ of the momenta of the atoms as a function of $t_\text{int}$. The triangles show the Fermi momentum $\Tilde{k}_\text{F}$, rescaled to the geometric mean of $\delta k_\text{x},\delta k_\text{y}$ at initial time $t_\text{int} = \SI{0}{\micro s}$.  The connecting lines serve as a guide to the eye. In the inset, the difference of $\delta k^2_\text{x}$ and $\delta k^2_\text{y}$ shows the build up of momentum anisotropy during the interacting expansion. The ideal hydrodynamic expansion (red line) and the  asymptotic long \textcolor{black}{time} limit derived from the real space data (red dashed line) provide a reference. 
    All error bars show the \SI{95}{\percent} confidence interval, determined using a bootstrapping technique.}
    \label{fig:expansion}
\end{figure*}

\subsection*{Experimental system and observables } 

We work with a mesoscopic system of ultracold fermionic $^6$Li atoms in the ground state of a potential created by two optical traps. The first one confines the atoms strongly in the vertical direction, creating an effectively two-dimensional (2D) system. The second trap (OT) confines the atoms in the horizontal plane, in an anisotropic, effectively harmonic confinement  with trap frequencies $(\omega _\text{x}, \omega _\text{y})/(2\pi) =  (\num{1280(3)}, \num{3384(7)})\unit{Hz}$ (see Methods).  We prepare a discrete many-body quantum state, composed of N spin up and N spin down atoms (denoted N+N) in the ground state, utilizing a technique developed in previous works \cite{Serwane_2011, Bayha_2020}.

\textcolor{black}{Our measurements are performed in a regime in which a separation of scales, traditionally underlying a  hydrodynamic description is not given.} The typical length scales of our system are on the order of the harmonic oscillator length, which is given by $l ^\text{x,y} _\text{HO} = \sqrt{\hbar / m \omega _\text{x,y} }  \approx (\num{1.1},\num{0.7}) \unit{\micro m}$, where $m$ is the mass of a $^6$Li atom. We estimate the typical interparticle spacing from the peak density $n_0 = (k_\text{F}^0)^2/(4\pi)$ of the non-interacting system, with the Fermi wave vector defined as $k^\text{0}_\text{F} = \sqrt{2mE_\text{F}}/\hbar$. Here  $E_\text{F} $ is the Fermi energy of the non-interacting system, determined by the highest filled energy level of the OT (see Methods). The mean interparticle spacing is  $1/\sqrt{n_0} \approx \SI{1.3}{\micro m}$. These length scales are estimated for the non-interacting system, but are on the same order of magnitude in the interacting case. \textcolor{black}{Assuming a kinetic description,}  the unitary limit would constrain the minimum mean free path to be on the order of the interparticle spacing. 

The strength of the attractive interactions can be tuned using the magnetic Feshbach resonance \cite{Zuern_2013}. It is quantified by the dimensionless interaction parameter $\text{ln}(k^\text{0}_\text{F} a_\text{2D})$, that relates the initial interparticle spacing (proportional to the inverse of the Fermi wave vector $k^\text{0}_\text{F}$) to the 2D scattering length $a_\text{2D}$ \cite{Shlyapnikov2001,Materials}. 

After preparing the system, we remove the horizontal confinement, while keeping the vertical 2D confinement. We let the atoms expand for an interacting expansion time $t_{\text{int}}$. At $t_{\text{int}}$, we instantaneously switch off interactions by a two-photon Raman transition \cite{Holten_2022}. Subsequently, we apply matterwave magnification techniques, to image either the momenta \cite{Holten_2022} or the  positions \cite{Brandstetter_2024} of the atoms at $t_\text{int}$. For the longest interacting expansion time ($t_\text{int} = \SI{9}{ms}$), the system has expanded enough for the atoms to be resolvable without matterwave magnification. 

We make use of a fluorescence imaging scheme to obtain single atom and spin resolved images \cite{Bergschneider_2018}. Each image represents a projection of the wave function on $N+N$ positions or momenta. We obtain the 2D density from approximately 1000 images for each setting (see Methods).  \textcolor{black}{To quantify the widths of the 2D densities, we calculate the root mean square (rms) $\delta$ of the momenta $k$ and positions $r$ over all atoms and images.}

\subsection*{Observing elliptic flow}

We investigate the evolution of the 2D density profiles in real and momentum space for a system of $5+5$ atoms and initial interaction parameter $\text{ln} (k^0_\text{F} a_\text{2D}) = \num{1.15}$. The obtained density profiles, superimposed with a randomly chosen single image of both spin states (black and white points), are shown in Fig. 1a-c (real space) and Fig. 1e-g (momentum space).

In real space (see Fig. 1a-d), the system starts from an elliptical density distribution with $\delta r_\text{x} > \delta r_\text{y}$  and expands anisotropically, with a stronger acceleration along the initially tightly confined direction. 
After $t_\text{int} \gtrsim \SI{250}{\micro s}$, one observes the characteristic inversion of the initial aspect ratio. Note that after long interacting expansion times we observe molecules (see Fig. 1c). 

In order to rule out single particle dynamics stemming from the Heisenberg uncertainty principle, we additionally measure the evolution of the momentum distribution (see Fig. 1e-h). In contrast to the one-body case, the system is initially isotropic, as expected in the many-body limit of a degenerate Fermi gas \cite{Pitaevskii_2016}. During the expansion, we observe two main effects in momentum space: the width in both directions decrease and an anisotropy builds up. Both processes subside at $t_\text{int} \approx \SI{160}{\micro s}$.

The increase of anisotropy arises from a redistribution of momenta driven by interactions. This observation is reminiscent of the dynamics of fluids and \textcolor{black}{Fermi gases} \cite{Schafer:2009zjw}, which exhibit a fast redistribution of momenta followed by the inversion of the initial aspect ratio. Therefore, we interpret our result as the signature of an emergent dynamics driven by a pressure-gradient-type force.

The initial width of the momentum distribution is determined by the Fermi momentum $k_\text{F} = \sqrt{4 \pi n} $ (see Fig. 1e dashed circle), obtained from the peak real space density $n(t_\text{int}) = N /(2 \pi \delta r_\text{x} \delta r_\text{y})$. To compare the Fermi momentum to the momentum space rms widths $\delta k_\text{x,y}$, we rescale $k_\text{F}$ to the geometric mean of $\delta k_\text{x}$ and $\delta k_\text{y}$ at the initial time $t_\text{int} = \SI{0}{\micro s}$ (denoted $\Tilde{k}_\text{F}$, shown in Fig 1h). As the system expands in real space the density drops and the Fermi momentum decreases accordingly (see Fig. 1e-g dashed circle and Fig. 1h). For short interacting expansion times, \textcolor{black}{$t_\text{int} \lesssim \SI{75}{\micro s}$}, the rescaled Fermi momentum $\Tilde{k}_\text{F}$ remains comparable to $\delta k _\text{x,y}$. For longer times, $\Tilde{k}_\text{F}$ is smaller than both momentum space widths. The Fermi momentum is comparable to $\delta k_\text{x,y}$ up to the time \textcolor{black}{where the redistribution of momenta subsides, indicating that this process only occurs in an interacting degenerate Fermi gas.}

In order to understand \textcolor{black}{which key features of the many-body system still persist in the few body limit, we compare our experimental results to the corresponding many-body system described by superfluid hydrodynamics}. The measured real space density at $t_\text{int}=\SI{0}{\micro s}$ (Fig. 1a) is used as an initial condition for the ideal hydrodynamic expansion, which is simulated by solving the continuity and Euler equation (see Methods). The pressure as a function of density is required as an additional input for the calculations. It is obtained from the equation of state (EOS) of the corresponding many-body system at low temperature, as determined in experiments on two-dimensional macroscopic Fermi gases \cite{2014PhRvL.112d5301M}. The comparison between the simulated real space evolution and the experimental data is shown in Fig. 1d.   

\textcolor{black}{While a fluid description does not directly describe particle momenta without additional assumptions, it is possible to relate fluid density and velocity to experimentally observed difference of momentum widths, }
\begin{equation}
    \label{eq:diff}
    \frac{m}{2N\hbar^2} \int_{\bf x} \rho(t, \mathbf{x}) \left[ v_y^2(t, \mathbf{x}) - v_x^2(t, \mathbf{x}) \right] = (\delta k_y)^2 - (\delta k_x)^2\, .
\end{equation}
This is achieved by relating the momentum flux in the fluid description (interacting atoms) to the one of freely streaming particles (non-interacting atoms)  (see Methods). This relation allows us to determine the  asymptotic limit of the squared momentum differences from the real space data.
We extract the difference in velocity fields from the simulation and compare it to the difference \textcolor{black}{of the squared} momentum widths $(\delta k _ \text{y} ^2 - \delta k _ \text{x} ^2)$, see inset Fig 1h.

\textcolor{black}{Experimentally, we observe a faster expansion of $\delta r_\text{x,y}$ compared to the ideal hydrodynamic simulation, however, the evolution of the aspect ratio matches the many-body prediction. Complementary, we also see qualitative agreement of the measured difference of the squared momentum widths and the simulation. An additional stress term, beyond the ideal fluid description used to model the many-body limit, could explain the deviation in the absolute sizes without altering the evolution of the aspect ratio. Such a stress term could stem from finite-size effects, that arise as the density changes faster than the mean interparticle spacing, breaking the validity of the Thomas-Fermi approximation \cite{Weizsaecker_1935, Pitaevskii_2016}. This can be taken into account by adding a density gradient dependent term -- known as the von-Weizsäcker term \cite{Weizsaecker_1935,Salasnich_2013} -- to the initial kinetic energy. }

\begin{figure}
    \centering
	\includegraphics{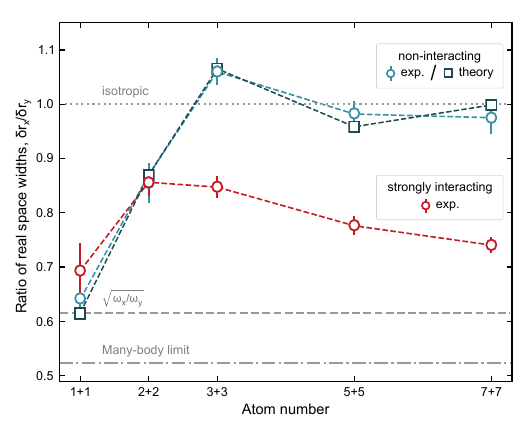}
    \caption{\textcolor{black}{\textbf {Emergence of interaction driven elliptic flow of few fermionic atoms.} }
    The anisotropy of the real space density after an expansion time of $\SI{9}{ms}$ is quantified by the ratio of the real space rms widths in $x$ and $y$ direction. The non-interacting system (blue dots experimental data, blue squares analytical calculation) evolves from the one-body limit, i.e. the inverse anisotropy of the harmonic oscillator ($\sqrt{\omega_x/\omega_y}$, dashed line) to the many-body limit, where we observe the build up of an isotropic (dotted line) Fermi surface. The strongly interacting system, characterized by $\text{ln}(k^0_\text{F} a_\text{2D}) = 1.15$ (red dots), starts from the same one-body limit. For the interacting system, we observe an inversion of the aspect ratio at all atom numbers, with a significant deviation from the non-interacting system \textcolor{black}{starting from 3+3 atoms}.
    The dashed dotted line shows the numerically calculated final aspect ratio in the many-body limit. The connecting lines serve as a guide to the eye. All error bars show the \SI{95}{\percent} confidence interval, determined using a bootstrapping technique.}
    \label{fig:atom_number}
\end{figure}

\subsection*{Building a fluid atom by atom}

\textcolor{black}{We study the emergence of elliptic flow for increasing particle numbers starting from a one-body system \cite{Floerchinger:2022qem}.} The dependency of the aspect ratio $\delta r_\text{x}/\delta r_\text{y}$ after a long expansion time on the atom number is displayed in Fig 2. We compare a strongly interacting system  ($\text{ln} (k^0_\text{F} a_\text{2D}) = 1.15$, red points) to the non-interacting system ($\text{ln} (k^0_\text{F} a_\text{2D}) \rightarrow \infty $, blue points) after an expansion time of \SI{9}{ms}. 

First, we investigate the non-interacting case. After a non-interacting expansion the final positions are determined by the initial momenta. In the non-interacting system of 1+1 atoms, we observe an inversion of the initial real space aspect ratio ($\delta r_\text{x}(\SI{0}{ms})/\delta r_\text{y}(\SI{0}{ms}) > 1$). The momentum space profile of 1+1 non-interacting atoms corresponds to that of a single atom in a harmonic oscillator. A single atom in the ground state has a momentum anisotropy given by the inverse anisotropy of the harmonic oscillator ($\sqrt{\omega_x/\omega_y}) $, which causes the inversion. For larger atom numbers, the non-interacting system becomes increasingly symmetric ($\delta r_\text{x}(\SI{9}{ms})/\delta r_\text{y}(\SI{9}{ms}) = 1$), as a round Fermi surface emerges in momentum space.  The measured aspect ratio agrees well with the analytical solution of the anisotropic harmonic oscillator ground state  (see Methods).

In the case of strongly interacting atoms $(\text{ln} (k^0_\text{F} a_\text{2D}) = 1.15)$ we see an inversion of the initial aspect ratio for all atom numbers.  \textcolor{black}{For 1+1 and 2+2 atoms, the interacting system behaves as the non-interacting case within the error of the data. 
However, starting at 3+3 atoms, we observe significant deviations, signalling an interaction driven inversion of the aspect ratio.} 

In the many-body limit, the local density approximation is applicable and the initial density distribution is described by the Thomas-Fermi (TF) approximation \cite{Stringari_2008}, whereby the spatial dependence of the chemical potential becomes determined by the shape of the confining harmonic potential. 
In the many-body limit, a prediction of $\delta r_\text{x}/\delta r_\text{y} \rightarrow \num{0.52}$ can be obtained from solving the \textcolor{black}{hydrostatic} equations using the TF density profile and the pressure determined by the many-body EOS as starting conditions (see Fig. 2) (see Methods). \textcolor{black}{The final aspect ratio approaches the many body limit with increasing atom numbers, indicating an emergent behaviour of the system.}

\begin{figure}
    \centering
	\includegraphics{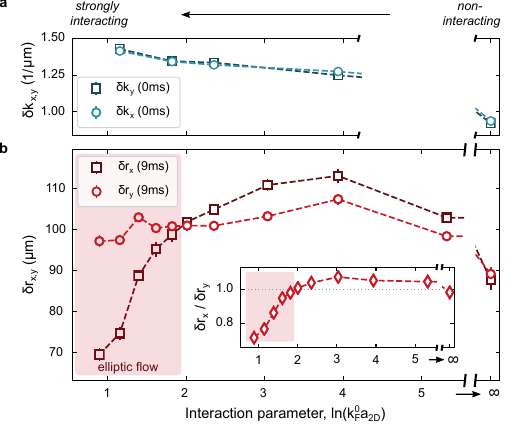}
    \caption{\textcolor{black}{\textbf{Interaction driven elliptic flow.}}
    We show the rms widths in momentum $\delta k_\text{x,y}(\SI{0}{ms})$ and real space $\delta r_\text{x,y}(\SI{9}{ms})$ for different interaction strengths. 
    \textbf{a} The initial momentum distribution remains isotropic for all interaction \textcolor{black}{strengths}, as expected for a degenerate Fermi gas and increases for stronger interactions.
    \textbf{b} The real space rms widths after $t_\text{int} = \SI{9}{ms}$ show the characteristic inversion of the aspect ratio for strong interactions. In the weakly interacting regime, the initial real space anisotropy is preserved. In the non-interacting limit it becomes isotropic, reflecting the initial momentum space profile. \textcolor{black}{The inset shows the ratio of the rms widths ${\delta r_\text{x} / \delta r_\text{y}}$.}
    The connecting lines serve as a guide to the eye. All error bars show the \SI{95}{\percent} confidence interval, determined using a bootstrapping technique.}
    \label{fig:interactions}
\end{figure}

\subsection*{Tuning interactions}

In order to investigate the influence of interactions on the expansion, we measure $\delta k_\text{x,y}$ at $t_\text{int} =\SI{0}{ms}$ and $\delta r_\text{x,y}$ after $t_\text{int} =\SI{9}{ms}$ for $5+5$ atoms at various interaction strengths. The momentum space data is shown in Fig. 3a, the real space data in Fig. 3b.

As expected for a degenerate Fermi gas, the initial momentum distribution is isotropic ($\delta k_\text{x}(\SI{0}{ms}) = \delta k_\text{y}(\SI{0}{ms})$)  for all interaction strengths. Stronger interactions yield a monotonous increase in $\delta k_\text{x,y}(\SI{0}{ms})$, as the size of the cloud in real space decreases due to the attractive interactions.

We identify two regimes for the interacting expansion, the strongly ($\text{ln} (k^0_\text{F} a_\text{2D}) \lesssim 2$) and the weakly ($\text{ln} (k^0_\text{F} a_\text{2D}) \gtrsim 2 $) interacting  regime . For strong interactions, we observe elliptic flow. The final real space distribution  becomes increasingly anisotropic with higher interaction strength. Additionally, the rms width along both directions reduces, with a stronger decrease of $\delta r_\text{x}$.

In the weakly interacting regime, the final real space widths $\delta r_\text{x,y}(\SI{9}{ms})$ do not reflect the shape of the initial momentum profile. Instead, the initial real space anisotropy is conserved ($\delta r_\text{x}(\SI{0}{ms}) >  \delta r_\text{y}(\SI{0}{ms}))$ \textcolor{black}{--  similar behaviour has been predicted for the collisionless normal Fermi gas where it originates from mean-field interactions~\cite{Menotti2002}}. This effect decreases with weaker interactions, and is not present in the non-interacting limit $\text{ln} (k^0_\text{F} a_\text{2D}) \rightarrow \infty$.

\subsection*{Conclusion and Outlook}

\textcolor{black}{We observe elliptic flow in a mesoscopic quantum gas of $^{6}$Li atoms. We show that the inversion of the aspect ratio is an interaction driven effect, and that it emerges with increasing atom number. It occurs in a regime where the established criteria for a hydrodynamic description do not hold, as in our experiment, the system size, the interparticle spacing, and the mean free path are not separable. }

Our experimental observations \textcolor{black}{link to the observation of collectivity in systems with small number of constituents, such as integrable systems~\cite{Malvania2021} or high-energy nuclear collisions~\cite{Nagle:2018nvi,Schenke:2021mxx}.} Furthermore, in analogy to the expansion of the quark-gluon plasma (QGP) the transition of our system from strongly interacting to free streaming is accompanied by the formation of bound states -- hadrons in the case of QGP and molecules in the case of cold atoms. Our experiment may thus relate to the chemical freeze out of the QGP \cite{Andronic:2017pug}. 

\textcolor{black}{To gain further insight into the mechanism behind the observed behaviour, we will investigate its connection to the emergence of superfluidity.} This has already been seen in helium nanodroplets with tens of atoms \cite{Grebenev_1998}. In our system, we can search for signatures of superfluidity by studying the rotational properties, and beyond that exotic strongly correlated quantum states \cite{Palm_2020}. 

%%%%%%%%%%%%%%%%%%%%%%%%%%%%%%%%%%%%%%%%%%%%%%%%%%%%%%%%%%%%%%%%%%%%%%%%%%%%%%%%%
%					     Bibliography                          		   			%
%%%%%%%%%%%%%%%%%%%%%%%%%%%%%%%%%%%%%%%%%%%%%%%%%%%%%%%%%%%%%%%%%%%%%%%%%%%%%%%%%

\bibliography{hydro.bib}
\bibliographystyle{naturemag.bst}
\setcounter{figure}{0}
\setcounter{equation}{0}
\renewcommand{\figurename}{Extended Data Figure}
\cleardoublepage
\newpage
\section*{Methods}

\paragraph*{\textbf{Preparation Sequence}}
A detailed description of the preparation scheme can be found in \cite{Bayha_2020}. We utilize the same steps to deterministically prepare stable ground state configurations of up to 7+7 atoms in a 2D harmonic oscillator. During the experimental sequence we use the hyperfine states of the $^2$S$_\frac{1}{2}$ Lithium ground state. The states are labelled in ascending order of energies $\ket{1}-\ket{6}$. 
First, the $^6$Li atoms are laser cooled by a Zeeman slower and in a magneto optical trap (MOT). From the MOT they are transferred into a red-detuned crossed beam optical dipole trap (CODT), where we utilize a sequence of radio frequency pulses to obtain a balanced mixture of atoms in the hyperfine states $\ket{1}$ and $\ket{3}$.

After evaporating in the CODT, we load the sample in a tightly focused vertical optical tweezer (OT). In the OT we evaporate further and make use of the spilling technique described in \cite{Serwane_2011} to end up with a sample of roughly 30 atoms where all states up to the Fermi surface are occupied with high probability.

To prepare the anisotropic 2D sample, we perform a continuous crossover to the quasi-2D regime by ramping up the power of a vertical optical lattice (2D-OT) with trap frequencies of $\omega _z / 2 \pi =  \SI{7432(3)}{Hz}$ and $\omega^\text{opt} _r / 2 \pi =  \SI{19.1(1)}{Hz}$. Simultaneously we decrease the radial confinement of the OT and adiabatically change its round shape into an elliptic one. The manipulation of the confinement and the shape of the tweezer is controlled by a spatial light modulator in the Fourier plane. Our final trap is an anisotropic 2D harmonic oscillator with trap frequencies of $\omega _x / 2 \pi = \SI{1280(1)}{Hz}$, $\omega _y / 2 \pi =  \SI{3384(7)}{Hz}$ and $\omega _z / 2 \pi =  \SI{7432(3)}{Hz}$ - see Extended Data Fig. 1.

By using the spilling technique introduced in \cite{Bayha_2020}, we prepare the 2D ground state of up to seven atoms per spin state deterministically. The anisotropy of our final trap leads to a different level structure compared to the isotropic harmonic oscillator described in \cite{Bayha_2020} (see Extended Data Fig. 1 b,c). With filled harmonic oscillator shells, we can prepare stable ground states of 1+1, 2+2, 3+3, 5+5 and 7+7 atoms with high fidelities. 

The interactions are described by the effective 2D scattering length $a_\text{2D}$, which is given by \cite{Shlyapnikov2001}
\begin{equation}
    a_\text{2D} = \ell_z \sqrt{\frac{\pi}{0.905}}\exp(-\ell_z/a_\text{3D}\sqrt{\pi/2}),
\end{equation}
where $\ell_z$ is the harmonic oscillator length in the vertical direction and $a_\text{3D}$ is the scattering length in a three dimensional (3D) system. The 3D scattering length between two different hyperfine states can be tuned via the magnetic offset field using a Feshbach resonance \cite{Zuern_2013}.

\begin{figure}
    \centering
	\includegraphics{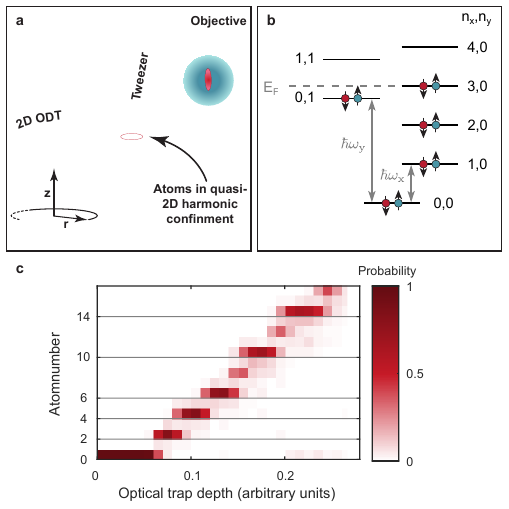}
    \caption{
    \textbf{Initial system} We prepare our initial system in a 2D harmonic oscillator. \textbf{a} The confinement is given by the overlap of a 2D light sheet (2D OT) and a vertical optical tweezer (OT) with elliptical beamshape. \textbf{b} The resulting level scheme of the non-interacting system is characterized by two quantum numbers (n$_\text{x}$, n$_\text{y}$). Shown is a closed shell configuration of 5+5 atoms. The Fermi energy is the energy of the highest occupied level. \textbf{c} By lowering the trap depth, the particles leave the trap shell by shell. Counting the number of atoms as a function of the optical trap depth reveals the level structure of the system. For the weakly-interacting system we find, corresponding to B, filled shells of 1+1, 2+2, 3+3, 5+5 and 7+7 atoms.      
    \label{som:fig:Stufenplot}
    }
\end{figure}

\paragraph*{\textbf{Interacting expansion}}
After preparing the groundstate in the OT, we turn the horizontal confinement off and let the atoms expand in a single plane of the radially symmetric 2D-ODT. We quench off the interactions at a given time $t_\text{int}$ after the release from the OT, by driving a Raman transition from the $\ket{1}-\ket{3}$ mixture to the almost non-interacting $\ket{1}-\ket{4}$ mixture. For the $\ket{1}-\ket{4}$ mixture, the \textcolor{black}{scattering} length is equal to that of the singlet with $a_\text{s}/a_\text{0} = \num{47(3)}$ \cite{Abraham_1996}. The Raman transition is performed on timescales on the order of \SI{300}{ns}. We have previously shown that this timescale is fast enough to preserve correlations in the system \cite{Holten_2022}. 

In the radial symmetric 2D plane, the radial confinement of the atoms is given by the gaussian optical trap and the harmonic magnetic trap. The optical trap frequency is $\omega^\text{opt} _r / 2 \pi =  \SI{19.1(1)}{Hz}$. The magnetic trap frequency depends on the magnetic field $B$ (given in \unit{G}) and is given by $\omega^\text{mag} _r / 2 \pi =  \SI{12.15}{Hz} \cdot \sqrt{\frac{B}{700}}$. 

For short interacting expansion times, the effect of the radial external potential is negligible. Compared to a free time of flight, the final position differs by less than $1\,\%$. For long interacting expansion times, the potential becomes relevant (deviation $\approx 25\%$), and also depends on the magnetic field. In the analysis we can account for this by solving the equation of motion in the combined trap numerically. 

\paragraph*{\textbf{Matterwave magnification}}
We utilize matterwave magnification techniques to obtain either the position or the momenta of the atoms after short interacting expansion times of up to $\SI{350}{\micro s}$. A detailed description for the momentum space measurements can be found in \cite{Holten_2022} and in \cite{Brandstetter_2024} for real space. 

To magnify the wavefunction, we switch off interactions at $t = t_\text{int}$. The switch-off of interactions is followed by the magnification of the wavefunction. To extract the momenta of the atoms, we let the atoms evolve for $t_\text{tof} = \SI{9}{ms}$ in the potential given by the combination of the 2D OT and the magnetic trap. During this expansion the cloud size increases by a factor of $\approx 50$, allowing us to resolve single atoms. Additionally this allows us to map the final positions of the atoms onto their momenta at $t=t_\text{int}$.  \cite{Holten_2022}

To image the positions of the particles at $t=t_\text{int}$, we utilize the matterwave magnification scheme described in \cite{Asteria_2021,Brandstetter_2024}. The interaction switch off is followed by an expansion in an optical trap with a trap frequency of $\omega_\text{exp}/2\pi = \SI{947}{Hz}$ for a quarter period. This maps the initial positions $x(t_\text{int})$ onto the momenta after the expansion in the optical trap. This is then followed by the same expansion sequence described for momentum space above. The ratio of trap frequencies allows us to magnify the initial positions by a factor of $\approx 42$. This allows us to resolve individual particles and measure their position after the matterwave magnification, which can be directly mapped back to their position at $t=t_\text{int}$ \cite{Brandstetter_2024}.

\paragraph*{\textbf{Imaging}}
A detailed description of the imaging technique can be found in \cite{Bergschneider_2018, Holten_2022}. The imaging protocol depends on whether we apply one of the matterwave magnification techniques. 

The exact imaging protocol when using matterwave magnification can be found in \cite{Holten_2022}. As discussed above, we utilize the non-interacting $\ket{1} - \ket{4}$ mixture for matterwave magnification. Due to technical reasons, the magnetic field is jumped to \SI{750}{G} after the interaction switch-off. During the non-interacting expansion time, we perform two subsequent Landau Zener passages from $\ket{1} \rightarrow \ket{2} \rightarrow \ket{3}$, as $\ket{3}$ has a closed imaging transition. We then image the atoms in $\ket{3}$. After this first image we perform a Landau Zeener passage from state $\ket{4} \rightarrow \ket{3}$ and again take an image of the atoms in $\ket{3}$. As states $\ket{3}$ and $\ket{4}$ are separated by $1.9\,\text{GHz}$, off resonant scattering is highly suppressed. 

For measurements without a matterwave magnification protocol, the atoms remain in the $\ket{1} - \ket{3}$ mixture. As the Landau Zeener passages described above take $\approx \SI{8}{ms}$, it is not possible to shift the atoms in $\ket{1}$ to $\ket{3}$, after taking the first image. The imaging transition for $\ket{1}$ is not closed and there is a finite possibility for the atom to decay into a dark state. For imaging atoms in $\ket{3}$, we achieve imaging fidelities of $\approx 98\%$, while the detection fidelity for atoms in $\ket{1}$ is only $\approx 93\%$. In addition, we observe a significantly higher number of off-resonant scattering events compared to the $\ket{1}$ -$\ket{4}$ mixture,  due to the smaller detuning of the imaging frequencies. To circumvent this problem, we first image the atoms in $\ket{3}$. These images are then neither affected by off resonant scattering, nor by the finite possibility to decay into a dark state. The atoms in state $\ket{1}$ are also imaged, but the positions are not utilized for the calculation of the rms width due to the abovementioned limitations. 

\paragraph*{\textbf{Non-interacting data}}

The single-particle wavefunction of the two-dimensional harmonic oscillator is given by

\begin{equation}
\begin{split}
    \Psi _{n,m} (x,y) = & \frac{1}{ \sqrt{ 2^n l^\text{x}_\text{HO} \pi ^{1/2}  n! } }  \text{e}^{ -x^2 / 2 {l^\text{x}_\text{HO}} ^2 } \mathcal{H}_n (x/l^\text{x}_\text{HO}) \\
    & \frac{1}{ \sqrt{ 2^m l^\text{y}_\text{HO} \pi ^{1/2} m!} } \text{e}^{ -y^2 / 2 {l^\text{y}_\text{HO}} ^2 } \mathcal{H}_m (y/l^\text{y}_\text{HO}),
\end{split}
\end{equation}

with the harmonic oscillator length in x-, and y-direction $l^\text{x}_\text{HO}$ and $l^\text{y}_\text{HO}$, respectively, and the Hermite polynomials of $n$th order $\mathcal{H}_\text{n}$. 

Exemplary, we show the experimental data and the theoretical calculation of the 5-particle probability density in Extended Data Fig. 2 and Extended Data Fig. 3.

\begin{figure*}
    \centering
	\includegraphics{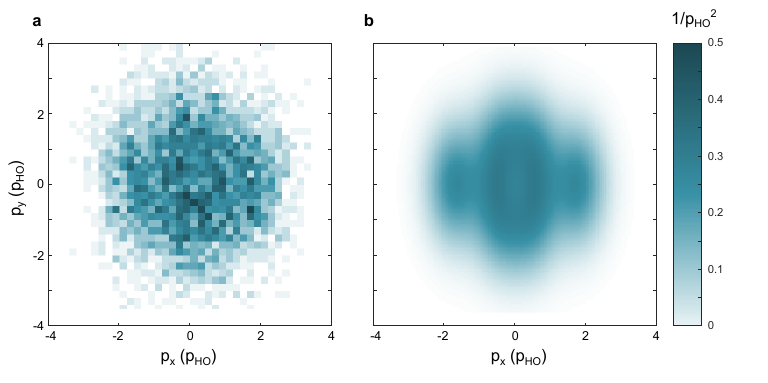}
    \caption{
    \textbf{Momentum space density of 5+5 non-interacting atoms.} 
    Comparison of the experimental and theoretical (\textbf{a} and \textbf{b}, respectively) momentum space density of 5+5 non-interacting atoms in our elliptical trap. The pixel size is given by the pixel size of our camera.
}
    \label{som:fig:5+5atoms_568G}
\end{figure*}

\begin{figure*}
    \centering
	\includegraphics{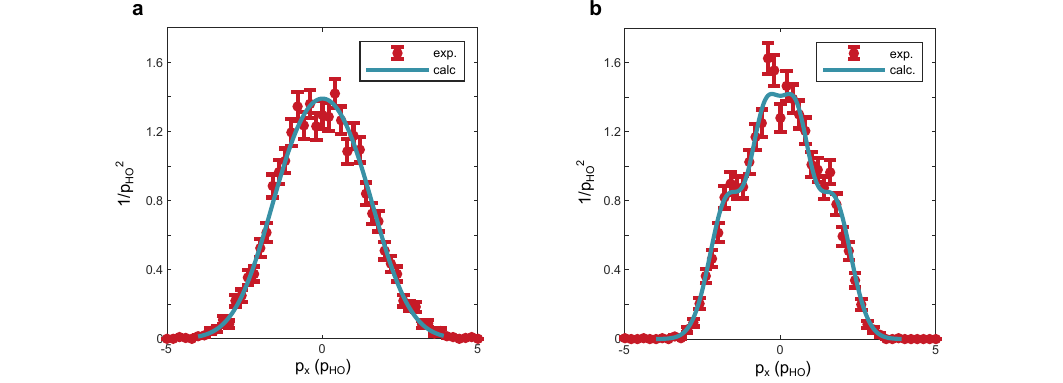}
    \caption{
    \textbf{Integrated momentum space density of 5+5 non-interacting atoms.} The measured and the calculated momentum density gets integrated along y- and x-direction (\textbf{a} and \textbf{b}, respectively). In the errors, the theoretical curve coincides with the experimental curves, showing that the non-interacting system is in the ground state of the harmonic oscillator potential.   
    }
    \label{som:fig:2+2atoms_568G}
\end{figure*}

\paragraph*{\textbf{Hydrodynamic simulations}}

The first step in the hydrodynamic modeling is the construction of the initial condition for the expansion, i.e., the mass density, $\rho$, at $t_{\rm int}=0$. 
\textcolor{black}{Although we lack theoretical guidance for the density of a strongly-interacting mesoscopic system, we note that a simple two-dimensional generalized Gaussian distribution of the type:
\begin{equation}
    \rho({\bf x},t_{\rm int}) \propto \exp \left ( - \left(\frac{|x|}{a_x}\right)^{b_x} \right ) \exp \left ( - \left(\frac{|y|}{a_y}\right)^{b_y} \right ),
\end{equation}
with $a_x=\SI{2.21}{\micro m}$ , $b_x=3$, $a_y=\SI{1.04}{\micro m}$, $b_y=2$, provides a very good description of the measured density in Fig.~1a.  The density is further normalized to satisfy:}
\begin{equation}
\label{eq:mtot}
    \int d^2{\bf x}~ \rho({\bf x},t_{\rm int}) = 2mN,
\end{equation}
where $m$ is the lithium mass. 

This initial density profile is then evolved in time according to ideal hydrodynamics. This amounts to solving equations for the conservation of mass, i.e., the fact that integral in Eq.~(\ref{eq:mtot}) is constant as a function of time, and the conservation of momentum \cite{Landau_1987}, respectively,
\begin{equation}
\begin{split}
\label{eq:ih}
    \partial_t \rho + \nabla \cdot (\rho {\bf v}) &= 0 \, , \\
  \rho   (\partial_t + {\bf v} \cdot \nabla) {\bf v} &= -\nabla P \, ,
\end{split}
\end{equation}
where ${\bf v}=(v_x,v_y)$ is the fluid velocity vector. This set of equations is closed by use of a thermodynamic equation of state that relates pressure, $P$, and density. To obtain the pressure, we match our system to the appropriate many-body limit of an interacting Fermi gas at zero temperature in two dimensions. The variation of $P$ as a function of the interaction strength parameter, $\eta = \ln (k_F a_{2D})$, 
is quantified as a deviation from the pressure of the ideal Fermi gas \cite{Levinsen2015}
\begin{equation}
   P_{\rm ideal} \equiv P (\eta\rightarrow\infty) = 
   \frac{\pi \hbar^2}{2 m^3} \rho^2 .
\end{equation}
In the range of $\eta$ relevant to our experimental setup, roughly $\eta<1.2$, an exponential correction to the ideal gas pressure accurately captures the result for $P(\eta)$ obtained 
from experimental measurements in equilibrium two-dimensional Fermi gases at low temperature \cite{2014PhRvL.112d5301M}. 
From the fit of the experimental data in the region of interest, $-1.7 < \eta < 1.2$, we obtain the correction
\begin{equation}
\label{eq:Pfit}
    P (\eta) = \alpha e^{\beta \eta} P_{\rm ideal},
\end{equation}
%with $\alpha=\num{0.277(8)}$ and $\beta=\num{0.53\pm0.02}$.} 
with $\alpha=\num{0.216\pm0.008}$ and $\beta=\num{0.67\pm0.05}$.
This allows us to stick to a polytropic-type EOS,
\begin{equation}
    P \propto \rho^{2+\beta/2}.
\end{equation}
Full solutions of the hydrodynamic equations are finally obtained by means of the compressible hydrodynamic solver of the \texttt{pyro} simulation framework \cite{pyro}. We compare to experimental data the temporal dependence of the dispersion of the system in real space,
\begin{equation}
    \langle \delta r_x^2 \rangle (t_{\rm int}) = \frac{\int d^2{\bf x}~ \rho({\bf x},t_{\rm int}) x^2}{\int d^2{\bf x} ~ \rho({\bf x},t_{\rm int})},
\end{equation}
and analogously for $\langle \delta r_y^2 \rangle (t_{\rm int})$.

\paragraph*{\textbf{Momentum space analysis}}

Hereafter we denote by $v_j$ the component of the velocity vector ${\bf v}$, where $j$ is either $x$ or $y$.
In the conservation law for mass in Eq.~(\ref{eq:ih}) the quantity $\mathscr{P}_{j} = \rho v_j$ corresponds to the mass current, or flux density, which defines the fluid velocity. In a nonrelativistic system, the mass current corresponds also to the momentum density. Hence, the local momentum conservation law is $\partial_t \mathscr{P}_k + \partial_{j}\mathscr{P}_{jk}=0$, where the symmetric tensor $\mathscr{P}_{jk}$ is the momentum flux density, with the subscript $k$ also labeling either $x$ or $y$. Fluid dynamics is based on an expansion around local thermal equilibrium in terms of gradients of the fields that characterize equilibrium states. The leading order truncation corresponds to ideal hydrodynamics, for which we have
\begin{equation}
\label{eq:Pjk}
    \mathscr{P}_{jk}(t,\mathbf{x}) = \rho(t,\mathbf{x}) v_j(t,\mathbf{x}) v_k(t,\mathbf{x}) + P(t, \mathbf{x}) \delta_{jk},
\end{equation}
where $P$ is the pressure and $\delta_{jk}$ is the Kronecker symbol. 

Experimentally, particle momenta are determined following an instantaneous switch-off of interactions. In a kinetic description, the momentum flux density of a non-interacting system with a phase space distribution $f(t,\mathbf{x},\mathbf{p})=dN/d^2 x d^2 p$  involves moments of the momentum distribution,
\begin{equation}
\label{eq:Pjkf}
    \mathscr{P}_{jk}(t, \mathbf{x}) = \int d^2 p \left\{\frac{p_j p_k}{m} f(t,\mathbf{x},\mathbf{p}) \right\}.
\end{equation}

The question is now whether the sudden change in interaction strength enables us to connect the interacting system to the non-interacting one by matching Eq.~(\ref{eq:Pjk}) and Eq.~(\ref{eq:Pjkf}) at $t_{\rm int}$.

The instantaneous interaction switch-off is not expected to change $\rho$ or $\textbf{v}$ because they are defined through conserved quantities, but it is expected to change the pressure term, $P$. In the simplest scenario, it would change from the pressure associated with the density in the interacting equation of state, to the one associated with a non-interacting equation of state, though non-equilibrium corrections beyond that should also be expected. 
To avoid assumptions about the dynamics of the isotropic pressure term, one can thus eliminate such a contribution by studying the difference in Eq.~(\ref{eq:diff}). 
This leads to a solid prediction of the hydrodynamic framework for quantities defined in momentum space.

 As discussed in the context of high-energy nuclear collisions \cite{Bhalerao:2005mm}, the build-up of momentum in a fluid occurs approximately over a time scale $\tau  = R/c_s$, where $R$ is the system size and $c_s$ is the speed of sound. In our case the system size is $R\sim \SI{1}{\micro m}$, while from the many-body EOS the speed of sound at the center of the cloud is of order $c_s = \sqrt{dP/d\rho} \sim \SI{10}{\micro m/ms}$. Consequently, from a hydrodynamic viewpoint the time scale of the build-up of momentum anisotropy in our system is of order $\tau \sim \SI{0.1}{ms}$. This is consistent with the trends  shown in Fig.~1H. 

\paragraph*{\textbf{Thomas-Fermi model}}

From a given equation of state one can obtain a hydrostatic solution in a trap. This is equivalent to the Thomas-Fermi approximation, where the chemical potential is replaced by the space-dependent quantity
\begin{equation}
    \mu(x) = \mu_0 - V(x) \, ,
\end{equation}
wherever the density is non-vanishing.
At constant (in our case zero) temperature we can recover from this and the equation of state $P(\rho)$ the density profile $\rho(x)$ by making use of $\dd{P} = n \dd{\mu}$. 
This holds as long as there are no relevant derivative corrections through finite size effects.
Here we observe deviations of the initial particle density from the Thomas-Fermi prediction that can hence be interpreted as a sign that finite size corrections play an important role in our system.
Performing hydrodynamic simulations with the fitted many-body EOS of Eq.~(\ref{eq:Pfit}) and starting with Thomas-Fermi profiles matched to different particle numbers, we find that the baseline aspect ratio at $t_{\rm int}=\SI{9}{ms}$ shown in Fig. 2 is independent of $N$. The origin of this finding as well as corrections to the Thomas-Fermi profile arising from the von-Weizsäcker term \cite{Weizsaecker_1935} will be addressed in a separate study.

\color{black}

\paragraph*{\textbf{Ballistic expansion}}

\begin{figure}
    \centering
	\includegraphics{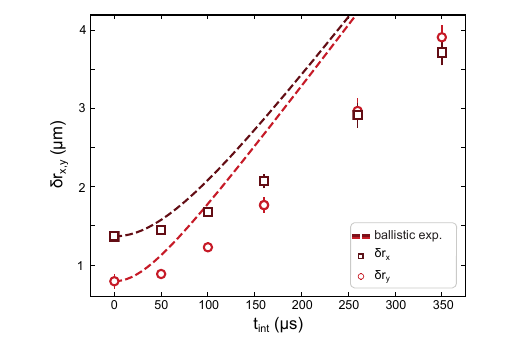}
    \caption{\textbf {Ballistic expansion.} Root mean square of the atom positions $\delta r _\text{x,y}$ as a function of $t_\text{int}$. The dashed lines show the expected ballistic expansion, the solid lines mark the ideal hydrodynamic evolution of the corresponding many-body system. 
    }
    \label{fig: ballistic_expansion}
\end{figure}

We compare the ballistic expansion to the experimental data in Extended Data Fig. 4.  The ballistic expansion is calculated using the time evolution of the position operator $\hat{x}$ to calculate $\langle \hat{x}^2(t)\rangle$. We assume the wavefunction to be Gaussian, such that $\langle \hat{x}(0) \hat{p}(0) \rangle + \langle \hat{p}(0) \hat{x}(0) \rangle$ vanishes. We plot 
\begin{equation}
    \sqrt{\langle \hat{x}^2(t) \rangle} = \sqrt{\cos^2{(\omega_{\text{r}} t)}\langle \hat{x}^2(0)\rangle + \sin^2{(\omega_{\text{r}} t)} \frac{\langle \hat{p}^2(0)\rangle}{m^2 \omega_{\text{r}}^2}}
\end{equation}
in the 2D ODT with trap frequency $\omega_{\text{r}}$ and $m$ being the Lithium mass. $\langle \hat{x}^2 \rangle$ and $\langle \hat{p}^2 \rangle$ are taken from the measurements. Note that $\langle \hat{x} \rangle$ and $\langle \hat{p} \rangle$ are zero. 
\color{black}
%\bibliography{hydro.bib}
%\bibliographystyle{Science.bst}

%%%%%%%%%%%%%%%%%%%%%%%%%%%%%%%%%%%%%%%%%%%%%%%%%%%%%%%%%%%%%%%%%%%%%%%%%%%%%%%%%
%					     Acknowledgements and Contributions   		   			%
%%%%%%%%%%%%%%%%%%%%%%%%%%%%%%%%%%%%%%%%%%%%%%%%%%%%%%%%%%%%%%%%%%%%%%%%%%%%%%%%%

\paragraph*{Data availability}
The data that support the findings of this study are available from the
corresponding authors upon reasonable request. Source data are provided with this paper.
\subsection*{Acknowledgments}
We gratefully acknowledge insightful discussions with Tilman Enss,  Aleksas Mazeliauskas, and Jean-Yves Ollitrault.
This work has been supported by the Heidelberg Center for Quantum Dynamics, the DFG Collaborative Research Centre SFB 1225 (ISOQUANT), the DFG project DFG FL 736/3-1 (NEQFluids), the Germany’s Excellence Strategy EXC2181/1-390900948 (Heidelberg Excellence Cluster STRUCTURES) and the European Union’s Horizon 2020 research and innovation program under grant agreements No.~817482 (PASQuanS) and  No.~725636 (ERC QuStA). This work has been partially financed by the Baden-Württemberg Stiftung.

\paragraph*{Author Contributions}
S.B. and P.L. contributed equally to this work. S.B.,\ P.L,\ C.H.\ and S.J conceived the experiment. S.B.\ and C.H.\ performed the measurements. P.L.,\ S.B.,\ and C.H.\  analyzed the data. S.J.\ supervised the experimental part of the project. S.F.,\ G.G.,\ and L.H.H. set up and ran the hydrodynamic simulations. S.B. and P.L.\ wrote the manuscript with input from all authors. All authors contributed to the discussion of the results.

\paragraph*{Competing Interest}
The authors declare no competing interests.
\paragraph*{Correspondence and requests for materials}
should be addressed to S.B. (brandstetter@physi.uni-heidelberg.de) and P.L. (lunt@physi.uni-heidelberg.de)

%%%%%%%%%%%%%%%%%%%%%%%%%%%%%%%%%%%%%%%%%%%%%%%%%%%%%%%%%%%%%%%%%%%%%%%%%%%%%%%%%
%					     Acknowledgements and Contributions   		   			%
%%%%%%%%%%%%%%%%%%%%%%%%%%%%%%%%%%%%%%%%%%%%%%%%%%%%%%%%%%%%%%%%%%%%%%%%%%%%%%%%%

\end{document}